# RIO EPICS device support application case study on an ion source control system (ISHP)

Diego Sanz [a], Mariano Ruiz [a,*], Mikel Eguiraun [b], Iñigo Arredondo [c], Inari Badillo [b], Josu Jugo [b], Jesús Vega [d], Rodrigo Castro [d]

[a] UPM – Universidad Politécnica de Madrid, Madrid, Spain
[b] Department of Electricity and Electronic, Faculty of Science and Technology, University of Basque Country, Bilbao, Spain
[c] ESS Bilbao Consortium, Zamudio, Spain
[d] Asociación EURATOM/CIEMAT, Madrid, Spain

## HIGHLIGHTS

- A use case example of RIO/FlexRIO design methodology is described.
- Ion source device is controlled and monitored by means EPICS IOCs.
- NIRIO EPICS device support demonstrates that is able to manage RIO devices.
- Easy and fast deployment is possible using RIO/FlexRIO design methodology using NIRIO-EDS.
- RIO/FlexRIO technology and EPICS are a good combination for support large scale experiments in fusion environments.

## ABSTRACT

Experimental Physics and Industrial Control System (EPICS) is a software tool that during last years has become relevant as a main framework to deploy distributed control systems in large scientific environments. At the moment, ESS Bilbao uses this middleware to perform the control of their Ion Source Hydrogen Positive (ISHP) project. The implementation of the control system was based on: PXI Real Time controllers using the LabVIEW-RT and LabVIEW-EPICS tools; and RIO devices based on Field-Programmable Gate Array (FPGA) technology. Intended to provide a full compliant EPICS IOCs for RIO devices and to avoid additional efforts on the system maintainability, a migration of the current system to a derivative Red Hat Linux (CentOS) environment has been conducted. This paper presents a real application case study for using the NIRIO EPICS device support (NIRIO-EDS) to give support to the ISHP. Although RIO FPGA configurations are particular solutions for ISHP performance, the NIRIO-EDS has permitted the control and monitoring of devices by applying a well-defined design methodology into the previous FPGA configuration for RIO/FlexRIO devices. This methodology has permitted a fast and easy deployment for the new robust, scalable and maintainable software to support RIO devices into the ISHP control architecture.

## 1. Introduction

Experimental Physics and Industrial Control Systems (EPICS) is a set of software tools developed for distributed control system implementations. This middleware has become relevant in recent years for supporting large scale scientific experiments, such as the Swiss Light Source (SLS) at Paul Scherrer Institute; the Laser Interferometer Gravitational-Wave Observatory (LIGO) operated by Caltech and MIT; The European Spallation Source project ESS-AB; or the International Thermonuclear Experimental Reactor (ITER).

The Ion Source Hydrogen Positive (ISHP) project consists of a highly versatile electron cyclotron resonance (ECR) ion source. It has been built with the aim of achieving several purposes. During the first stages, this project should provide a workbench to test accelerator-related technologies and validate in-house made developments, control systems and data acquisition (DAQ) electronics. Following this idea, the ECR ion source has to be adjustable in order to be able to support a diversity of experiments and equipment. It implies that its control system has to be very flexible to incorporate new hardware elements and permitting fast integration into the system for new experiments, tests, etc. [1,2]. These required features have been tested and favored in previous works [3,4] using LabVIEW EPICS; this is because they are the best candidates for interfacing a human operator with the entire system and specific hardware devices.

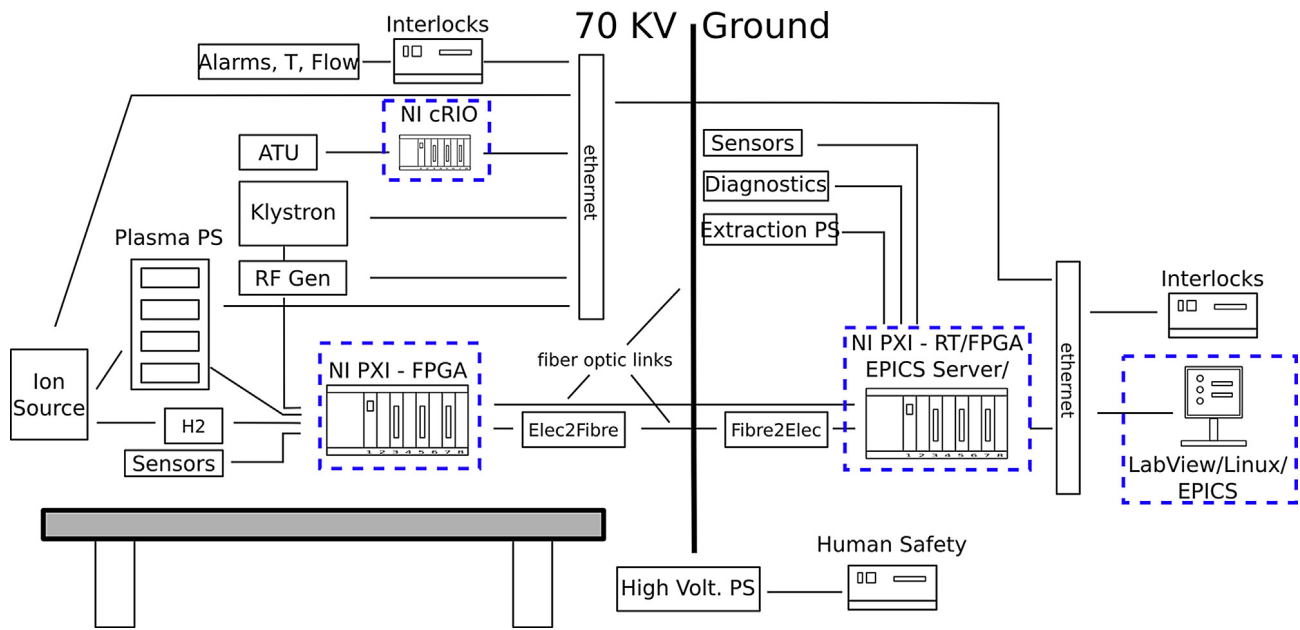

**Fig. 1.** Schematic structure of the ISHP control system architecture. The system is divided into the areas of high voltage platform and ground. The system is managed by two PXI chassis and one cRIO for the control tasks, as well as several PLCs in charge of safe operation.

The hardware technology used to carry out the ISHP control is mainly supported by PXI and cRIO based on Field-Programmable Gate Array (FPGA). This is not usual in large EPICS-based facilities, although there are some exceptions, like the Spallation Neutron Source (SNS) [5], which only makes use of this approach for the beam diagnostics. Similarly, but not based on EPICS, this solution is also extended for the diagnostics at ISIS [6].

In the present case, the PXI chassis are equipped with reconfigurable input/output (RIO) devices and an additional cRIO device based on FPGA equipped with standard I/O DAQ cards. The software used for controlling all the devices was entirely based on the graphical programming language LabVIEW for real-time (RT) and the implementation of non-fully compliant EPICS IOC was based on the LabVIEW toolkit module.

Although LabVIEW offers a user-friendly and fast programming environment, IOC servers are not fully EPICS compliant. In addition, future integrations of the current control system into a full accelerator control environment will require robustness, compatibility and maintainability for the entire software architecture. This is the reason why the IOC server software for RIO devices is migrated to a Linux version. However, LabVIEW for FPGA is absolutely necessary for the FPGA hardware configuration.

To accomplish this task, a well-defined methodology has been used [7], which allows the implementation of intelligent data acquisition systems (I-DAQs) based on RIO/FlexRIO technology and integration into Linux EPICS environments. This methodology is supported by a RIO/FlexRIO EPICS device support (NIRIO-EDS) based on asynDriver technology, which reassigns all FPGA resources dynamically to the corresponding process variables (PVs). It also launches DMA threads according to the FPGA DMA channels configured to send the data acquired to the host RAM.

This is the first time that this methodology has been applied to a real and operative system. However, this methodology was previously defined to support I-DAQs implementation for ITER fast plant system controllers [4,7].

## 2. Hardware architecture description

An ECR ion source forms plasma by injecting RF microwaves into a vacuum chamber containing a gas and a proper magnetic field. Then, the ions of the plasma are extracted by means of an electrical potential difference. All of the equipment needed to generate the beam has to be on a high voltage platform in order to give the initial accelerating potential to the particles.

Fig. 1 shows a schematic of the control system for the ion source with the main elements. Due to the high voltage platform, the control system is divided in the ground and platform areas. The platform has the main elements required for plasma generation (RF power source, hydrogen, power supplies, etc.), and the ground includes the beam diagnostics and several power supplies among other devices. Several PLCs and safety PLCs ensure a safe operation of the experiment for both the personnel and the machine. Because the present work addresses the control aspects, the next paragraphs will only explain the details of the control network, specifically for the systems where the proposed EPICS driver has special importance. Additional information regarding the general control system of this project can be found in [1].

The main hardware for the ISHP control is based on PXI and cRIO devices. Two PXI chassis, one on the platform and the other one at the ground, are interconnected by a fiber optic MXI link to avoid the high voltage gap. The grounded one (NI PXIe-1065) is equipped with a PXIe controller (NI PXIe-8108), in which LabVIEW RT, with EPICS DSC tools, controls the devices in both chassis (the chassis in the platform NI PXI-1042Q). The set-up of the chassis is composed of two RIO PXI-7852R (one in the platform chassis and one on the ground) and several communication cards (PXI-8433/4). The serial cards are used to communicate with several motor drivers, while the FPGAs are in charge of managing some additional motors; implementing data acquisition; managing several power supplies; controlling the $H_2$ flow; and implementing the trigger sequence to the RF amplifier for the generation of the plasma. Complementarily, the cRIO handles all the signals of the RF system.

In the initial software implementation there is a LabVIEW IOC server that implements the software support to control and monitor RIO devices through the channel access (CA) EPICS protocol. The primary interface between the PXI controller and the EPICS control network is based on the tools provided by the LabVIEW DSC module. The interface also runs on the LabVIEW RT operating system in the PXI controller. The EPICS server implemented in the PXI was improved for integrating the current control system into a general one. For this purpose, the existing EPICS PVs are connected into the main control system and the full characteristics of an EPICS IOC are added in an external PC [3].

This LabVIEW IOC server is the software unit that is going to be replaced by the EPICS IOCs that is supported by the NIRIO-EDS mentioned above. In addition, an external computer is no longer required for completing EPICS features. This new software unit will require a Linux Red Hat derived version and the use of Linux kernel modules (nirio kernel driver). The following section addresses the migration procedure, which uses the novel design methodology based on NIRIO-EDS.

## 3. Migration of the software using a predefined design methodology

As mentioned before, RIO devices are FPGA based, and every RIO requires a particular hardware configuration according to its particular functionalities within a DAQ system. For the ISHP control, RIO devices require: I/O registers; FPGA to host data streaming through a DMA channel; and control logic to support the motor system control and the implementation of the triggering sequence. It seems obvious that the new Linux software would require specific codification to support the ISHP RIO devices. However, following the design methodology, NIRIO-EDS will be able to manage the specific implementation for the ISHP case without substantial modifications on both FPGAs.

The design methodology consists of the following established steps: scientific requirements defined by the scientist in charge; system specification, defined by a group of engineers; LabVIEW for RIO/FlexRIO implementation using a set of rules [8] and patterns developed by the programmer of the LabVIEW for FPGA code; auto-generation of records database and the st.cmd startup EPICS files; and finally the EPICS IOC execution.

In this particular case, in which the behavior of the RIO devices is perfectly defined for the previous operational control system [3], it is possible to start the design methodology from LabVIEW during the RIO implementation step.

RIO devices require to be configured under a set of mandatory design rules in order to be supported by the NRIO-EDS. As mentioned above, the stages of scientist specifications and system requirements can be avoided, but the FPGA configuration would require some modifications according with the design rules. This means that, although the previous RIO configuration had not been deployed using the methodology, it is possible to adapt the LabVIEW code without hard efforts to integrate it to a Linux EPICS environment.

NIRIO-EDS uses the RIO device as a black box with a set of I/O registers; a number of DMA channels for high throughput data transmission; and a profile that defines what the behavior of the RIO device is (see Fig. 2).

At this point, the design rules only permit to implement I-DAQ systems (coreDAQ profile) and intelligent image data acquisition systems (I-IMAQ) (coreIMAQ profile). In this case the coreDAQ profile has been chosen. This profile requires the implementation of some specific read registers into the FPGA, where the driver will obtain specific information from the hardware configured into the FPGA. This information is related to the number of analog inputs packaged by the DMA channel such as the clock frequency with which the sequential logic hardware will be fed, etc.

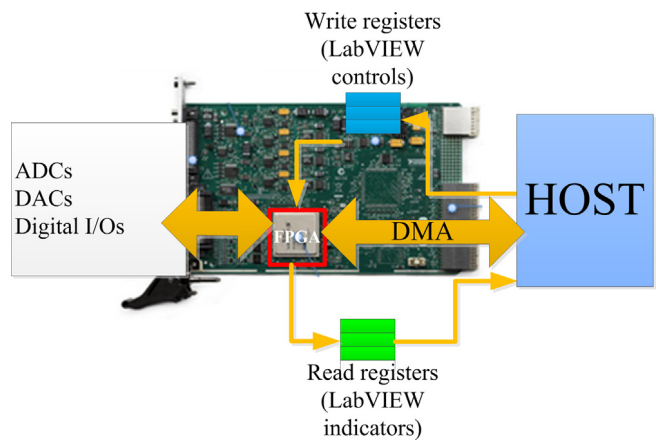

**Fig. 2.** RIO device is observed as a black box with I/O registers and DMA channels by the NIRIO-EDS.

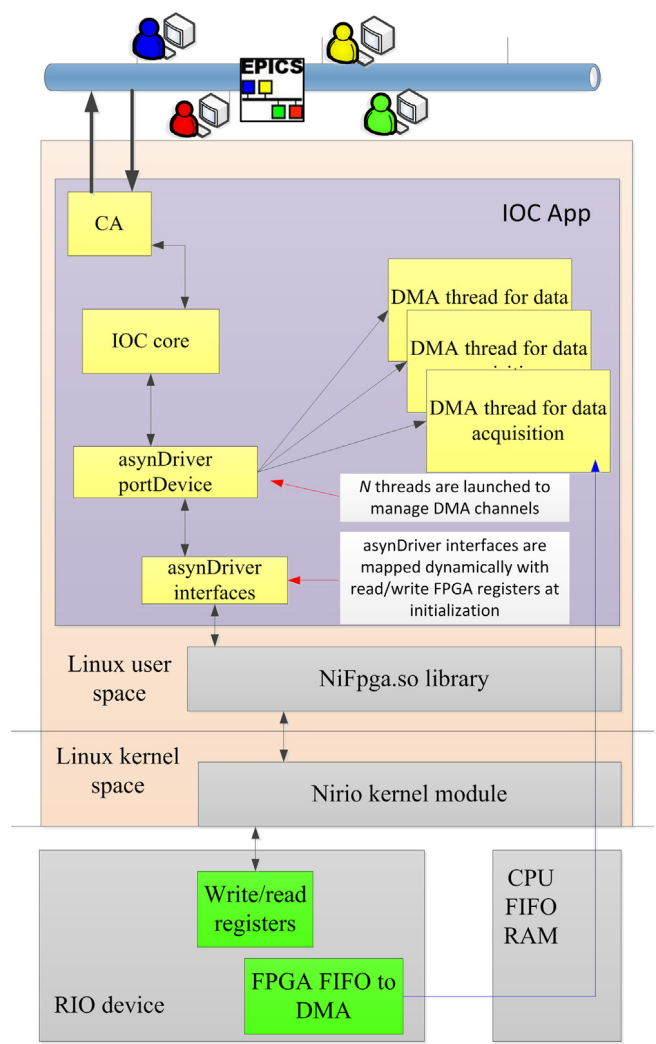

**Fig. 3.** The IOC EPICS software architecture based on the NIRIO-EDS, which interfaces all FPGA resources with the corresponding PVs.

Fig. 3 shows the software architecture that implements an IOC application based on the NIRIO-EDS. When the NIRIO-EDS is initialized, it launches the required threads to attend the DMA channels, which will unpack all the raw data acquired by the RIO device.

The steps that have been followed to reconfigure the RIO devices are as follows: adding all the base registers required by the NIRIO-EDS driver to identify the FPGA resources and the integrated functionalities; redefining all I/O FPGA register names according to the design rules. This permits to add both scalar integers, float I/O registers, and vector I/O registers. That is, changing the controls and indicators names in the LabVIEW code.

By describing correctly (following the design rules) how the FPGA packs all the acquired data to be sent to the host RAM, it is not necessary to modify the rest of the hardware described in the LabVIEW code. After the FPGA LabVIEW code is compiled, the bitfile-stream is obtained. By means of the LabVIEW C API generator tool and using the bitfile-stream, a file with all the mapped FPGA resources is generated. Both files will be used by the NRIO-EDS to configure the FPGA from the RIO device, to dynamically assign every PV from the record database to every FPGA I/O register and to launch $N$ DMA threads to retrieve all the data acquired.

Once all the FPGA resources (DMA channels, channels per DMA, etc.) and all the name modifications are registered on the auto-generation tool, the record database and the st.cmd EPICS start-up file will be generated. All these steps are summarized in Fig. 4.

## 4. Results and conclusions

This example applied to a real ion source experiment has demonstrated that the design methodology, defined by RIO/FlexRIO

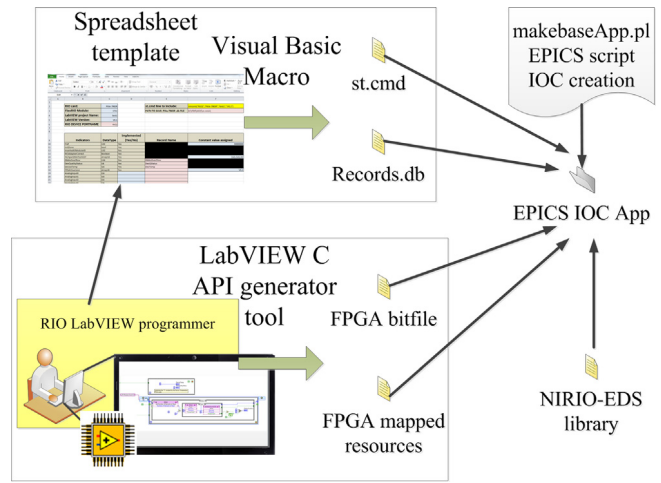

**Fig. 4.** Steps to configure a RIO device, to be supported by the NIRIO-EDS.

and supported by the NIRIO-EDS, has satisfied the migration and integration with a Linux software environment.

With short modifications to the LabVIEW code for FPGA in order to describe the FPGA behavior according to the design rules, it has been possible to integrate RIO/FlexRIO devices into EPICS environments, permitting distributed device control and monitoring.

In this particular case, the NIRIO-EDS supports most of the previous control features such as the data acquisition, the control logic for motor control, and the generation of the triggering sequence for the RF generator and the beam diagnostics acquisition, etc. All of the code and the features that were initially programmed in the FPGAs remain the same as before and functioning properly.

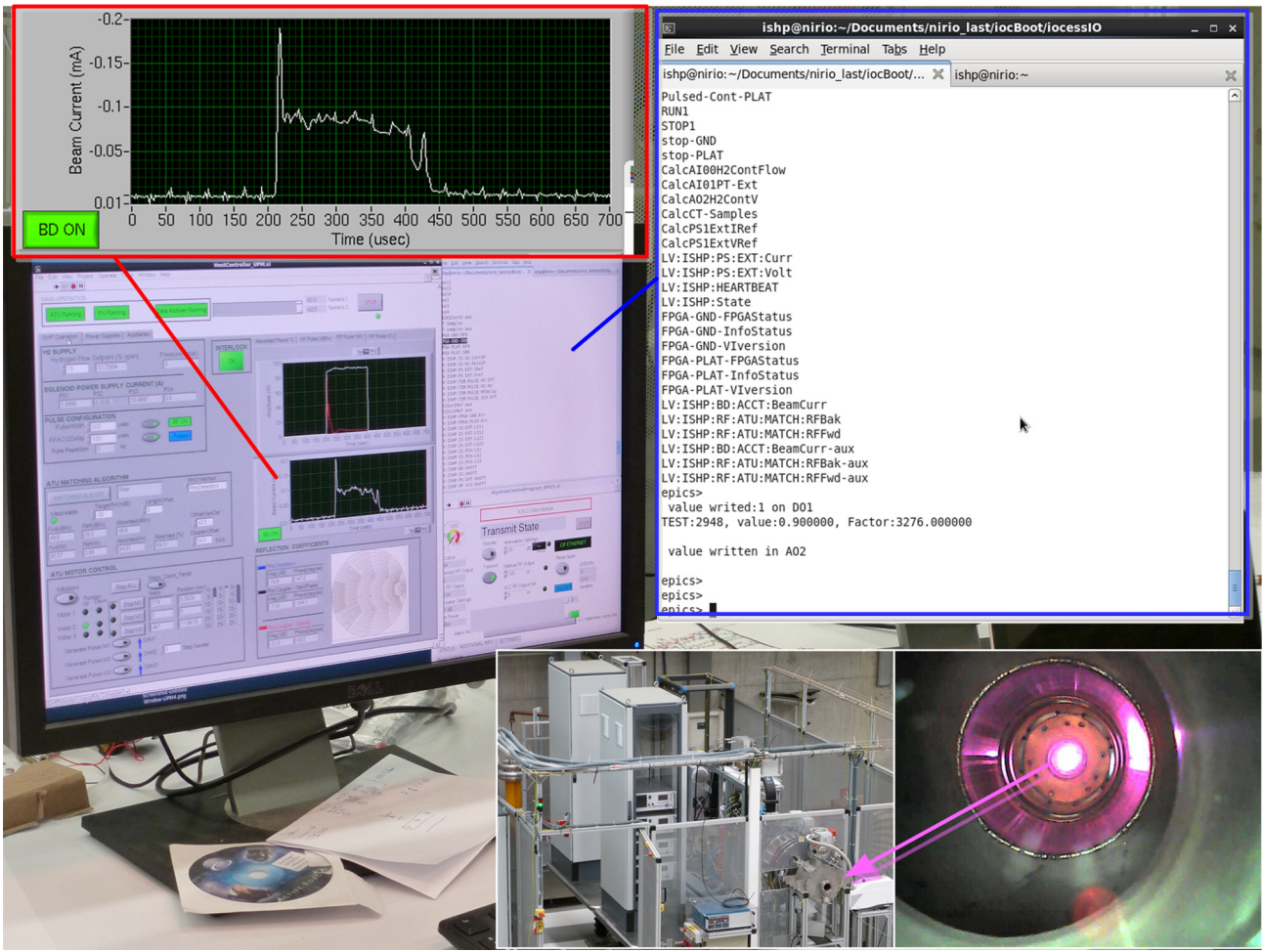

**Fig. 5.** The new EPICS IOCs implemented using RIO/FlexRIO design methodology supported by the NIRIO-EDS, working on a real case application for the ESS Bilbao ISHP.

Fig. 5 shows the ISHP during an experiment. The top left image shows a beam pulse acquired with an ACCT diagnostic; the mid-left screen shows the graphical user interface developed in LabVIEW for Linux and the interfacing EPICS for all of the communications; the terminal on the right shows the IOC shell; and finally the photographs at the bottom show an overview of the facility and an image of the plasma.

## References

[1] I. Arredondo, M. Eguiraun, J. Jugo, D. Piso, M. del Campo, T. Poggi, et al., Adjustable ECR ion source control system, in: Real time conference 2014, Nara, Japan, 2014.
[2] I. Arredondo, M. Eguiraun, D. Piso, M. del Campo, J. Feuchtwanger, G. Harper, et al., Sistema de Control de la fuente de iones ISHP, in: National Instruments case study, 2013.
[3] M. Eguiraun, J. Jugo, I. Arredondo, M. del Campo, J. Feuchtwanger, V. Etxebarria, et al., ISHN ion source control system. First steps toward an EPICS BASED ESS-Bilbao accelerator control system, IEEE Trans. Nucl. Sci. 60 (April (2)) (2013) 1280–1288.
[4] M. Ruiz, J. Vega, R. Castro, D. Sanz, J.M. López, G. de Arcas, et al., ITER Fast Plant System Controller prototype based on PXIe platform, Fusion Eng. Des. 87 (12) (2012) 2030–2035.
[5] C.D. Long, M.P. Stockli, T.V. Gorlov, B. Han, S.N. Murray, T.R. Pennisi, Control system for the Spallation Neutron Source H-source test facility Allison scanner, Rev. Sci. Instrum. (2010) 02B722.
[6] S.J. Payne, P.G. Barnes, G.M. Cross, A.H. Kershaw, N. Leach, A. Pertica, et al., Beam diagnostics at ISIS, 2008.
[7] D. Sanz, M. Ruiz, R. Castro, J. Vega, J.M. Lopez, E. Barrera, et al., Implementation of intelligent data acquisition systems for fusion experiments using EPICS and FlexRIO technology, IEEE Trans. Nucl. Sci. 60 (2013) 3446–3453.
[8] D. Sanz, Modeling and integration of intelligent data acquisition systems into instrumentation systems for fusion devices (Ph.D. dissertation), Univ. Politécnica de Madrid, Madrid, Spain, 2014.